\documentclass[twocolumn,letterpaper]{IEEEAerospaceCLS}  

\usepackage{amssymb}
\usepackage[]{graphicx}    
\newcommand{\ignore}[1]{}  

\usepackage[acronym,toc,nonumberlist,style=super,nogroupskip]{glossaries}
\newacronym{AbMCC}{AbC}{Angle-based Correlation}
\newacronym{ACE}{ACE}{Attitude Control Error}
\newacronym{ADCS}{ADCS}{Attitude Determination and Control System}
\newacronym{AKE}{AKE}{Attitude Knowledge Error}
\newacronym{PSF}{PSF}{Point Spread Function}
\newacronym{SEPP}{SEPP}{Satellite Experimental Processing Platform}
\newacronym{SNR}{SNR}{Signal-to-Noise Ratio}
\newacronym{TLE}{TLE}{Two-Line Elements}
\newcommand\abmcc{\gls{AbMCC}}
\newcommand\ADCS{\gls{ADCS}}
\newcommand\ACE{\gls{ACE}}
\newcommand\AKE{\gls{AKE}}
\newcommand\psf{\gls{PSF}}
\newcommand\sepp{\gls{SEPP}}
\newcommand\snr{\gls{SNR}}
\newcommand\tle{\gls{TLE}}
\usepackage{xcolor, soul}
\sethlcolor{yellow}

\begin{document}
\title{Refined Astrometry on Board a CubeSat}

\author{%
Boris Segret\\ 
CENSUS (1), LESIA Paris Observatory (2)\\
PSL Université Paris (3)\\
Boris.Segret@observatoiredeparis.psl.eu
\and 
Youssoupha Diaw\\
CENSUS (1), LESIA Paris Observatory (2)\\
PSL Université Paris (3)\\
Youssoupha.Diaw@observatoiredeparis.psl.eu
\and 
Valéry Lainey\\
IMCCE Paris Observatory (4)\\
PSL Université Paris (3), CNRS (5)\\
lainey@imcce.fr
\thanks{\footnotesize (2022 IEEE preprint)}              }

\maketitle

\thispagestyle{plain}
\pagestyle{plain}

\maketitle

\thispagestyle{plain}
\pagestyle{plain}

\begin{abstract}
\ Optical navigation on a CubeSat must rely on the best extraction of the directions of some beacons from on-board images. We present an experiment on OPS-SAT, a CubeSat of the European Space Agency (ESA), that will characterize an on-board algorithm to this aim, named \abmcc. OPS-SAT is a 3-unit CubeSat with an \ADCS\ and an imager that have proved their reliability with typical performance at CubeSat scale. We selected a few star-fields that all present enough visible stars within a 10$\,^\circ$ field of view. When our experiment is run, OPS-SAT is pointed to the most convenient star-field at that time. There, the star-field is imaged and subwindows are extracted from the image around the expected location of each star, based on the attitude-quaternion reported by the \ADCS. The \abmcc\ reconstructs the absolute direction of the central body, in principle unknown, which is the pointed known star in the experiment. The method intensively uses the quaternion algebra. The beacon location is first consolidated in the field of view with the \abmcc. Then, the field of view is finely positioned against the sky, again with the \abmcc. A covariance is associated with the found beacon direction. Our experiment with OPS-SAT manages the pointing and the imager, and processes the taken images. Then, it downloads the on-board computed absolute directions and their covariances, to be compared with the actual directions. After a campaign of intensive use of the experiment, the statistical performance of the algorithm will be established and compared to the on-board computed covariances. As a bonus, an assessment of OPS-SAT's inertial pointing stability will be available. The \abmcc\ can theoretically get rid of the \ACE\ of the platform and of the \AKE\ estimated by the \ADCS, and potentially converge to sub-arcsec measurements, whereas 15 to 90\,arcsec for \AKE\ are expected on CubeSats and imager pixel size ranges in 10 to 15\,arcsec. Results will be available from 2022, thus, instead, we illustrate here possible applications for commissioning and autonomous operations.
\end{abstract}

\tableofcontents

\subsubsection{Affiliations~}~
(1)~\emph{CENSUS}: CEnter for Nano-Satellites in Universe Sciences, l'Observatoire de Paris, 5 pl. Jules Janssen, 92195 Meudon Cedex, France.
(2)~\emph{LESIA}: Laboratoire d'Etudes Spatiales et d'Instrumentation en Astrophysique, Observatoire de Paris, 5 pl. Jules Janssen, 92195 Meudon Cedex, France.
(3)~\emph{PSL Université Paris}: Paris Sciences Lettres Université Paris, 60 rue Mazarine, 75006 Paris, France.
(4)~\emph{IMCCE}: Institut de M\'ecanique C\'eleste et de Calcul des Eph\'em\'erides, Observatoire de Paris, 61 av. Denfert Rochereau, 75014 Paris, France.
(5)~\emph{CNRS}: Centre National de la Recherche Scientifique, France.

\section{Introduction}
The navigation of deep space missions has always relied on images of beacons taken on board that, most of the time, are processed on ground. The derived optical data are aggregated with data of other kinds (e.g. ranging from radio-tracking techniques) to feed a process of orbit determination of the observer (the spacecraft itself), of the imaged beacon (e.g. a celestial body's ephemerides), or both. Eventually, the result is the basis to design the on-board operations for trajectory correction maneuver or some target tracking, or to expand the scientific data-set about the celestial body that served as a beacon. On ground, the techniques that extract the optical data from a space image, represent a broad field of astrometry as an input to derive scientific knowledge from the dynamics of celestial bodies (see \cite{Cooper2018} and references therein).

Whereas on-ground techniques are complex and can be secured by a human analysis, very few missions have derived optical data on board. The techniques used on board NASA's \textsc{Deep Space 1} were presented in 1997 as an experiment \cite{Riedel1997} and are summarized below. They were inherited from on-ground techniques, initially with manual measurements on the received pictures, for \textsc{Voyager 2} to Uranus (1986, \cite{Synnott1986}) and Neptune (1989, \cite{Riedel1990}), then for \textsc{Galileo} to asteroids Gaspra (1991, \cite{VAUGHAN2013}) and Ida (1993, \cite{Antreasian1994}). The centerfinding techniques were very efficient to produce directions at sub-pixel accuracy at a time when CCD was still a novel technology in space, with long exposure times and poor platform stability. The Single-Frame Mosaic technique consists in exposing the same celestial scene on multiple areas of the sensor within a single downloadable image: at each exposure, the platform slews, which applies the same pattern on the images of all celestial objects of the field of view (beacon of interest and neighbouring stars). The multiple exposure technique is intended to limit the volume of downloaded data and is  not detailed here. Then, the image produces by each long exposure serves as a filtering technique called ``box filtering'': the smeared image of each object produces a pattern within an area of pixels, or box, whose maximum size can be anticipated; the pattern in each box is the same and is also searched in the box where the celestial beacon of interest is expected; a convolution product of one box with all other boxes yields the shifts of the patterns from one another in different boxes. Then, a ``multiple cross-correlation'' is performed by taking each known star as the convolution kernel successively, which produces a set of measurements, weighted by the \snr\ of each box. Eventually, a least-squares method fits the measurements with the location of the beacon of interest at an expected accuracy better than 0.1 pixel. Although the centerfinding techniques proved to be efficient even with faint objects, the overall autonomous navigation experienced in \textsc{Deep Space 1} was not adopted operationally \cite{Rayman2000}. The astrometric reduction has mainly remained a task assigned to the ground segment of the space missions. It allows extensive algorithms and checks that are obviously worth being performed on ground, although it requires an important telecommunication budget.

The next section will present our \abmcc\ method in details. Section \ref{ID156} reminds ESA's OPS-SAT mission and our experiment ``Id.156'' in OPS-SAT. Section \ref{applications} presents our possible application and calls for further use with various contexts and hardware.

\section{Angle-based astrometric reduction}

The \abmcc\ method is a full process combining linear estimate optimization and quaternion algebra. After some context, we review the possible errors in the chain of measurements. Then, we compute an estimate of the beacon location and its covariance in the focal plane.  Then, we transform this image location into an absolute direction of the sky by reversing the gnomonic projection, which first requires a fine re-centering and rotation of the image against the sky.

\subsection{Context of use, notations}
The typical use of the presented \abmcc\ is the processing of an image that contains an unresolved foreground celestial body, the ``beacon'', and several stars compatible with the imager's sensitivity. Whether such stars are available is discussed below in the numerical applications of this section. Then, the beacon is assumed to be present within a maximum parallax $\sigma_\pi$ from a pre-computed direction, based on mission preparation work, that is not discussed here. The imager is pointed to the pre-computed direction of the beacon and puts it in a central region of the image, whose half-size is lower than $\sigma_\pi+\sigma_\textsf{ACE}$, where $\sigma_\textsf{ACE}$ is the attitude control error of the platform (including the imager, assumed to be fixed). In this region of the image, the platform's \ADCS\ informs about the estimated pointing with an accuracy $\sigma_\textsf{AKE}$. Typically, with nowadays CubeSat hardware, one can expect an imager with a field of view $\textsf{FoV} \sim 10^\circ$ and an \ADCS\ with $\sigma_\textsf{ACE} \sim 0.1^\circ$ and $\sigma_\textsf{AKE}$ in $15''$ to $90''$. Obviously, the beacon needs to be in the central region, which limits the maximum acceptable parallax to $\sigma_\pi + \sigma_\textsf{ACE} \ll \textsf{FoV}$.

A pre-processing step is to extract the location of the stars and the beacon from square subwindows of the image where they are expected, i.e. subwindows whose edge is $6\sigma_\textsf{AKE}$ for the stars and $\sigma_\pi+6\sigma_\textsf{AKE}$ for the beacon, centered at their expected locations. This extraction is made by a centerfinding algorithm, for instance but not necessarily a ``box filtering'' as mentioned in the introduction, or a least-squares fit of a point spread function, or even a simple photometric barycenter.

\begin{figure}[t]\centering
   \includegraphics[width=0.95\linewidth]{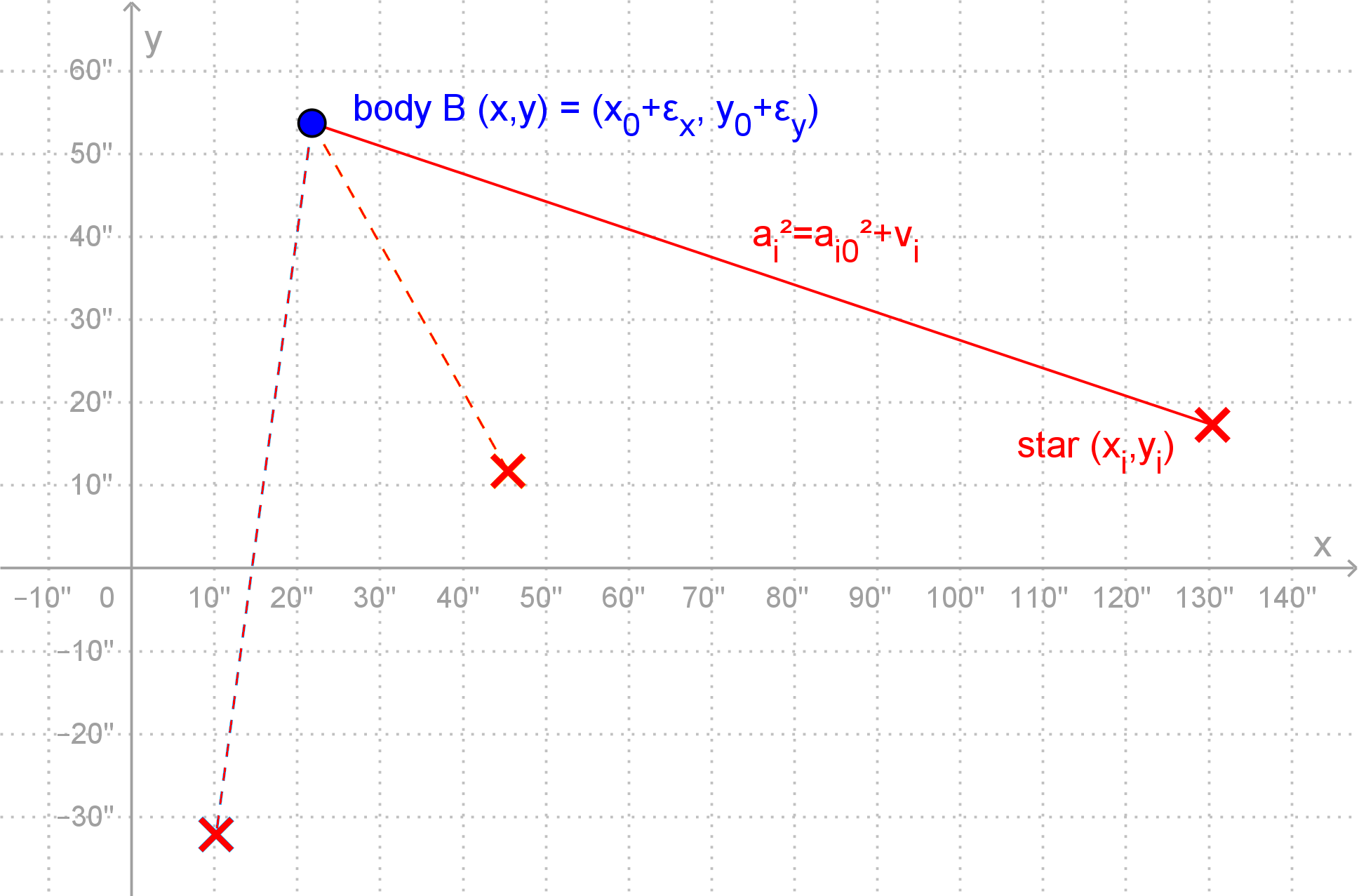}
   \caption{\label{fig:OTpfocal}Notations: ``B'' is the foreground beacon, ``X'' are images of known background stars.}
\end{figure}

As a result of the image pre-processing, the centerfinding yields raw locations on the image (Fig.\,\ref{fig:OTpfocal}): $B(x,y)$ for the beacon and $\{S_i(x_i,y_i),\ i=1\dots n^\star\}$  for the $n^\star$ known stars. The actual direction of the beacon corresponds to a location $(x_0,y_0)$ on the image. The actual directions of the stars in the image are noted $\{(x_{0i},y_{0i}),\ i=1\dots n^\star\}$ and correspond to known directions of the sky, as given by a catalog prepared on ground and stored on board.

\subsection{Review of the measurement errors}

The expected locations $(x_{0i},y_{0i})$ are the gnomonic projections of the stars on the reference axes of the imager, perpendicular to its line of sight $\vec z_\textsf{ADCS}$, as expected on the basis of the \ADCS\ information. In the absence of error, a bijection exists between the directions in right ascension and declination inside the cone toward the line of sight, with semi-angle $\pi/2$ and the coordinate axes $\{\vec x,\vec y\}_\textsf{ADCS}$ of the imager:

\[\begin{aligned}
\textbf{p}_g:~ \textsf{cone}(\vec z_\textsf{ADCS},\pi/2) \rightarrow~ & {\mathbb{R}}^2 \\
 (\textsf{RA}_i,\textsf{dec}_i) \mapsto~ & (x_{0i},y_{0i})_\textsf{ADCS}
\end{aligned}\]

The expression of $\textbf{p}_g$ or its reverse function are not trivial. They are discussed after the presentation of the \abmcc, in this section. Nevertheless, $(x_{0i},y_{0i})$ differ from the centerfinding results $S_i(x_i,y_i)$ due to the following possible errors, summarized in Table \ref{errors}:

\subsubsection{Attitude knowledge~}~
The attitude knowledge quaternion provided by the \ADCS\ will be updated just before and just after the exposure, then interpolated at the timestamp of the image. The information about jitter and axis rotation errors are not available, hence the interpolated attitude is associated with the uncertainty \AKE, which typically ranges within 15-90\,arcsec. The attitude control error, \ACE, is not at stake here as the subwindows are selected on the basis of the attitude knowledge.

\subsubsection{Non-uniformity of the pixels and pixel matrix~}~
Photometric dispersion is expected. Different centerfinding techniques, that are differently affected, are considered in the presented experiment, which is dedicated to the improvement of the centerfinding results, rather than the centerfinding itself. Hence, these errors exist but are not individually quantified here. The possible dispersion covers shot noise, read noise, dark current and its bias, pixel-to-pixel sensitivity, stray-light, discretization. Nevertheless, hot pixels should be neutralized in the centerfinding techniques (i.e. skipped in the computations) whenever they are known. As well, cosmic rays hitting a subwindow can introduce a temporary strong distortion in the centerfinding, unless multiple images are taken and compared to neutralize hitten pixels, meaning additional pre-processing. Irregularities may still exist in the shape of the pixel lines of a detector, and of the pixels in a line. They are expected lower than a fraction of an averaged pixel and would introduce a photometric bias in each pixel and, consequently, in the centerfinding process as well. The photometric quality of the centerfinding will serve to weight the relative trust in each $S_i(x_i,y_i)$, if positioning errors were equal for each $i$, based on the \snr, noted $\sigma_\textsf{SNR,i}$, used in the \abmcc.

Whereas $\sigma_\textsf{SNR,i}$ is a weighting factor for the quality of the images in the subwindows, the following errors affect the location of the imaged centers themselves.

\subsubsection{Star catalog accuracy and parallax~}~
Considering that the catalog used for the reference stars is up-to-date (or can be updated), the proper motion of a star is not considered here. The parallax of a star is also given in the catalogs. In the scope of the presented experiment, in Low Earth Orbit, the greatest parallax for visible stars is for Proxima Centauri at 0.77\,arcsec. Depending on the date of the year, the parallax can be included in the catalog for all stars with an accuracy better than 4.2\,milli-arcsec per day since the catalog's update. Say $\sigma_\textsf{catalog}$ the maximum error since the last update.

\subsubsection{Stellar aberration~}~
It affects the apparent direction of all objects in the field of view due to the velocity of the observer. The largest effect is in the direction perpendicular to the velocity, adding an apparent angle of $d\alpha=\arctan(v/c)$ in the forward direction, e.g. $d\alpha=20.6$\,arcsec with $v=30$\,km/s. Within an $\textsf{FoV}=10^\circ$, this angle decreases down to $d\alpha'=\cos(\textsf{FoV})d\alpha$, e.g. 20.3\,arcsec at the considered velocity. This is the maximum bias difference from one edge to the opposite edge of an image. Say $\sigma_\textsf{St.Ab.}$ the maximum error for the considered $\textsf{FoV}$.

\subsubsection{Optical aberrations~}~
Optical aberrations move and distort an off-axis image. Through a lens, they also depend on the wavelength. Typically, in imagers at CubeSat scale (pixels of $\sim 15$\,arcsec), it is likely to be expected that the projection of a star can shift by up to one or two pixels wrt to its gnomonic projection, depending on its color and its distance to the line of sight. Say $\sigma_\textsf{OPT,i}$ this error, with the index $i$ to remind the non-uniformity of the error.

\subsubsection{Misalignments of the imager~}~
The surface of the detector may be not perfectly perpendicular to, nor centered with the optical axis of the imager. A mathematical transformation can be introduced in the gnomonic projection onto the pixel matrix of the imager to correct for misalignments, usually performed during the imager commissioning, and based on calibration images taken in flight. However, such correction introduces third-order terms related to the distance to the optical axis, that expresses correlations between dimensions $x$ and $y$ of the pixel matrix. Although expected to be much lower than a fraction of a pixel, the experiment will be an opportunity to check for such aberrations with images of known star-fields. Say $\sigma_\textsf{AIT,i}$ this error, with the index $i$ to remind the non-uniformity of the error.

\begin{table}
\renewcommand{\arraystretch}{1.3}
\caption{\bf Expected errors for the experiment on OPS-SAT}
\label{errors}
\centering
\begin{tabular}{|l|c|c|}
\hline
\bfseries Type of error & \bfseries Variable & \bfseries Values \\
\hline\hline
Attitude knowledge error    & $\sigma_\textsf{AKE}$     & 15..90\,arcsec \\
Photometric dispersion      & $\sigma_\textsf{SNR,i}$   & $\geq 3$ \\
Star catalog, 1 month old   & $\sigma_\textsf{catalog}$ & 0.13\,arcsec \\
Stellar aberr. diff. within 10$^\circ$ & $\sigma_\textsf{St.Ab.}$ & 0.3\,arcsec \\
Optical aberrations         & $\sigma_\textsf{OPT,i}$ & 1..2 pixels \\
Detector misalignments      & $\sigma_\textsf{AIT,i}$ & 0.1 pixel \\
\hline
\end{tabular}
\end{table}

\subsection{Mathematical development}

The \abmcc\ is first used, here, to estimate the shift vector $\hat E = (\varepsilon_x,\varepsilon_y)^T$ from $(x_0,y_0)$ to $(x,y)$, along with a variance-covariance matrix (``covariance'' in short).

The \abmcc\ takes advantage of the fact that the angular distances from the beacon to the neighbouring stars are constrained by one another. The square of the true angular distance between the beacon and a given star is:
\begin{equation}\label{eq:ang_dist}
 a_{0i}^2=(x_0-x_{0i})^2+(y_0-y_{0i})^2 .
\end{equation}

An estimate of Eq.\ref{eq:ang_dist} is given by the measurements:
\begin{equation}\label{eq:ang_meas}
 a_i^2=(x-x_i)^2+(y-y_i)^2.
\end{equation}

For all $a_i^2$ of the same image, $(\varepsilon_x,\varepsilon_y)$ are parameters that can be fit to minimize the residuals $\nu_i = a_i^2 - a_{0i}^2$, obtained from the measurements $S_i(x_i,y_i)$ and $B(x,y)$. Then, replacing $x_0$ and $y_0$ by $(x-\varepsilon_x)$ and $(y-\varepsilon_y)$ in Eq.\,\ref{eq:ang_dist}, and developing $\nu_i$ at the first order, a linear system of equations is built:

\[\begin{array}{l}
\begin{pmatrix}\vdots \\ a_i^2-(x-x_{0i})^2-(y-y_{0i})^2 \\ \vdots\end{pmatrix} =\\
\begin{pmatrix}
\vdots & \vdots \\ -2(x-x_{0i}) & -2(y-y_{0i}) \\ \vdots & \vdots
\end{pmatrix}\begin{pmatrix}\varepsilon_x \\ \varepsilon_y\end{pmatrix}
+\begin{pmatrix}\vdots \\ \nu_i \\ \vdots\end{pmatrix}
\end{array}\]

or

\begin{equation}
\label{eq:A=CE+N}
A = CE + N,~\textsf{where}~N \sim {\mathcal{N}}(0,V),
\end{equation}

where the notation $\sim {\mathcal{N}}(0,V)$ defines the left-hand part as a zero-mean Gaussian random variable with covariance $V$. The system can be inverted by a weighted least-squares method, with a weight matrix $W=V^{-1}$. Then, an estimate for $E=(\varepsilon_x,\varepsilon_y)^T$ and its covariance are given as follows:

\begin{equation}
\label{eq:OTsol}\begin{aligned}
\widehat{E} &= (C^T W C)^{-1} C^T W A
\\
\textsf{Cov}(E) &= (C^T W C)^{-1}.
\end{aligned}
\end{equation}

The matrix $(C^TWC)$ has rank 2 and its elements are easy to compute provided $W$ is known. Their expression is developed in appendix. In a simplified approach to assess the diagonal matrix $V=\textsf{Diag}\{\textsf{Var}(\nu_i)\}$, that is discussed in the conclusions of this section, we first consider that the \snr\ is the same in all subwindows and that the geometric centerfinding errors for the stars and the beacon are dominated by a picture-wide uniform, zero-mean, Gaussian error, with standard deviation $\sigma_\textsf{in}$ expressed in arcsec:

\[\begin{aligned}
\sigma_\textsf{in}^2 \geqslant \sigma_\textsf{catalog}^2 + \sigma_\textsf{St.Ab.}^2 &+ \textsf{max}\{\sigma_\textsf{OPT,i}^2\}+\textsf{max}\{\sigma_\textsf{AIT,i}^2\}
\\
(x_i-x_{0i}) &\sim {\mathcal{N}}(0,\sigma_\textsf{in}^2)
\\
(y_i-y_{0i}) &\sim {\mathcal{N}}(0,\sigma_\textsf{in}^2)
\end{aligned}\]

Considering that $x_{0i}$ and $y_{0i}$ are fixed for a given picture (derived from a catalog), as well as $x_0$, $y_0$ and the parameters $(\varepsilon_x,\varepsilon_y)$, $\textsf{Var}(\nu_i)$ can be derived, as detailed in appendix, from Eq.\ref{eq:ang_dist} and \ref{eq:ang_meas}:

\begin{equation}\label{eq:OTopt}
\forall i \in\{1..n^\star\}, \textsf{Var}(\nu_i)=4 \sigma_\textsf{in}^4
\end{equation}
Hence, the expression of the rank $n^\star$ diagonal matrix $W=(4\sigma_\textsf{in}^4)^{-1} \, I_{n^\star}$. Although easy to compute in the general case, $\textsf{Cov}(E)=(C^TWC)^{-1}$ is not analytically simple. However, in an ideal case where the image of the beacon were at the barycenter of a homogeneous cloud of star images, the expression becomes simple. Noting $d$ the standard deviation for the coordinates $x_{0i}$ and $y_{0i}$ of the stars, and expecting a diagonal $2\times2$ matrix $\textsf{Cov}(E)$ in the form of $(\sigma_{\textsf{out}}^2)\,I_2$, we can show after some development (given in appendix) that:

\begin{equation}\label{eq:OTsigma}
\sigma_{\textsf{out}} = \frac{\sigma_\textsf{in}}{\sqrt{n^\star}}\left(\frac{\sigma_{\textsf{in}}}{d}\right).
\end{equation}

\subsection{Numerical applications}

As an indication for our experiment based on Table \ref{errors}, with 15\,arcsec pixels, $\sigma_\textsf{in}=30$\,arcsec (2 pixels) should be an upper bound, due to a risk of strong optical aberrations.

\begin{figure}[t]\centering
   \includegraphics[width=0.95\linewidth]{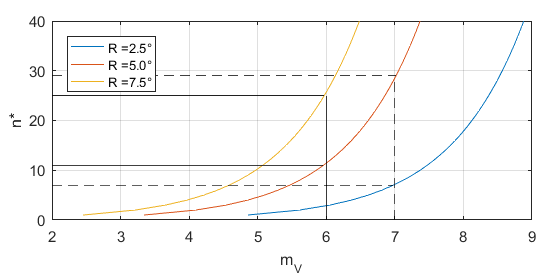}
   \caption{\label{fig:NbOfStars}Average number of stars $n^\star$ by visual magnitude $m_v$ in a conic field of view of half-angle R.}
\end{figure}

Eq.\,\ref{eq:OTsigma}, although in an ideal case, is promising with $d$ expected in hundreds of pixels and only a few stars: e.g. $\sigma_\textsf{in}=2$\,px, $d=100$\,px and $n^\star=5$ yields $\sigma_\textsf{out}=0.017$\,px or 0.27\,arcsec. Fig.\,\ref{fig:NbOfStars} shows an average number of stars per surface of the sky. It is built from a total number of 6\,000 stars brighter than the $6^\text{th}$ magnitude and considering that their number increases by a factor of 2.5 per magnitude. An average of 5 stars can be found in a conic field of view of half-angle 5$^\circ$ if the sensor's sensitivity is better than $m_v \sim 5.5$ whereas the sensitivity of CubeSat imagers is expected between $m_v$ 6 (i.e. 12 stars in average) and 7 (i.e. 28 stars in average). These considerations have justified the experiment of the \abmcc\ on board OPS-SAT, to study its sensitivity in real flight conditions, with real CubeSat hardware, non-uniform star-fields, strong pointing deviations (\ACE\ up to 0.1$^\circ$) and optical aberrations.

\subsection{Gnomonic projection $\textbf{p}_g$ and its reverse projection}

The \abmcc\ method is used a second time to fit the field of view with its projection on the sky. Indeed, the measurement $B(x,y)$ is performed in the pixel matrix, not on the sky. Hence, a direct use of the quaternion provided by the \ADCS\ to come back from $(x_0,y_0)=(x-\varepsilon_x, y-\varepsilon_y)$ to an absolute direction of the sky, using the reverse gnomonic projection of $(x,y)$, would re-introduce the error \AKE\ in the results.

After defining some notations, a first quaternion $q_1$ is computed with the \abmcc, that transforms from the \ADCS -reported attitude to the best estimate of the line of sight. Then, a second quaternion $q_2$ to rotate the field of view is computed by minimizing the cost function of a least-squares method.

\subsubsection{Additional notations~}~
Let's consider the quaternion $q_\textsf{IMG}$ to rotate from the satellite body-frame to the imager field of view, and $q_\textsf{ICRF}$ to rotate from the ICRF to the target. In mission preparation, the quaternion to orient the imager to the desired target is computed, say $\overline q$ and the resulting quaternion that is requested to the \ADCS\ is $q_\textsf{ICRF} = \overline q \otimes q_\textsf{IMG}$, with ``$\otimes$'' the product in the quaternion algebra (it is reminded in appendix to avoid some ambiguity). However, the \ADCS\ reports a different quaternion $q_\textsf{ADCS} \neq q_\textsf{ICRF}$ that corresponds to a quaternion $q \neq \overline q$ for the imager:

\begin{equation}\label{eq:q_initial}
q = q_\textsf{ADCS} \otimes q_\textsf{IMG}^\star,
\end{equation}

where $q_\textsf{IMG}^\star$ is the conjugate of $q_\textsf{IMG}$ (assuming only unit quaternions).

One can note that neither $q_\textsf{ADCS}$ nor $q_\textsf{ICRF}$ is the quaternion that brings to the field of view, noted $q_0$ that is to be determined. Considering a unit vector $\vec s(s_x,s_y,s_z)$ in ICRF coordinates, we note $R(\vec s)=(s_x',s_y',s_z')$ its coordinates in the frame of the imager. Its associated quaternion is given by:

\[
\left(0+R(\vec s)\right)= q \otimes (0+\vec s) \otimes q^\star.
\]

In the frame $(\vec i_\textsf{IMG},\vec j_\textsf{IMG})$ of the imager ($\vec k_\textsf{IMG}$ being the line of sight), whose focal length is $f_\textsf{IMG}$, the coordinates $(x,y)=\textbf{p}_g(\vec s)$ of the projection of $\vec s$ are:

\[\begin{aligned}
x &= f_\textsf{IMG} \, s_x'/s_z' \\
y &= f_\textsf{IMG} \, s_y'/s_z'
\end{aligned}\]

\subsubsection{Re-centering the field of view~}~
The measurements $S_i(x_i,y_i)$ on the image field of view of the cataloged stars are searched by the centerfinding pre-processing technique in regions centered at their \emph{expected} locations given by $q$, then noted $(x_{0i},y_{0i})_q$. Considering the line of sight as the beacon in the \abmcc\ method, i.e. $(x,y)=(0,0)$, Eq.\,\ref{eq:OTsol} returns a solution noted $\hat E_0=(\varepsilon_x,\varepsilon_y)_{(0,0)}$ as the best translation to fit $S_i(x_i,y_i)$, i.e. the rotation from $q$ to $q_0$  about an axis $\vec T$ in the plane of the field of view, perpendicular to this translation. In other terms, the beacon measured in $(0,0)$ corresponds to the direction $(x_0,y_0)_=(0,0)_{q_0}$ in such a way:

\[\begin{aligned}
(x_0,y_0)_q &= -(\varepsilon_x,\varepsilon_y)_{(0,0)} \\
 &= (0,0)_{q_0}
\end{aligned}\]

This solution defines a small rotation, hence a quaternion noted $q_1$, to re-center the field of view against the sky, from its \ADCS -reported quaternion. Hence, the quaternion to be used in the new gnomonic projection $\textbf{p}_g'$ is adapted from Eq.\,\ref{eq:q_initial} as follows:

\begin{equation}\label{eq:q_quote}\begin{aligned}
q' &= q_1 \otimes q \text{, where} \\
q_1 &= \cos \frac{\theta}{2} + \sin \frac{\theta}{2} \vec T / \Vert \vec T \Vert, \\
\text{with } \vec T &= \vec z_\textsf{ADCS} \times \textbf{p}_g^{-1}((-\varepsilon_x,-\varepsilon_y)_{(0,0)}) \\
\text{and } \theta &= \sqrt{\varepsilon_x^2+\varepsilon_y^2}.
\end{aligned}
\end{equation}

\subsubsection{Fine rotation of the field of view~}~
Solving the last degree of freedom to characterize the field of view against the sky consists in correcting its rotation error, assumed to be small. Still using the set of measurements $S_i(x_i,y_i)$ for the known stars in the picture (the values are unchanged, it is unnecessary to process a centerfinding again), a least-squares method is used to build a cost function $F(\omega)$ whose single parameter is the rotation angle about the line of sight of the new gnomonic projection $\textbf{p}_g'$ of the cataloged stars $\vec s_i$:

\begin{equation}\label{eq:F(w)}
F(\omega) = \sum_{i=1}^{n^\star} \Vert \overrightarrow{S_i \, \Omega(S_{0i})} \Vert ^2,
\end{equation}

where $\Omega(S_{0i})$ is the rotation by the angle $\omega$ about the center of the field of view of $\textbf{p}_g'(\vec s_i)$.

After some development at the second order (provided in appendix), the minimum of $F(\omega)$ is found for:
\begin{equation}\label{eq:omega}
\omega \textsf{ [rad]} = \frac{
\sum_{i=1}^{n^\star} (x_{0i} y_i - y_{0i} x_i) }{
\sum_{i=1}^{n^\star} (x_{0i} x_i + y_{0i} y_i) }.
\end{equation}

The value for $\omega$ in Eq.\,\ref{eq:omega} defines a new quaternion $q_2$ associated with the gnomonic projection $\textbf{p}_g"$, adapted from Eq.\,\ref{eq:q_quote}:
\begin{equation}\label{eq:q_dblquote}\begin{aligned}
q" &= q_2 \otimes q' = q_2 \otimes q_1 \otimes q \text{, where} \\
q_2 &= \cos \frac{\omega}{2} + \sin \frac{\omega}{2} \vec z_0 / \Vert \vec z_0 \Vert, \\
\text{with } \vec z_0 &= (\textbf{p}_g')^{-1}(0,0).
\end{aligned}
\end{equation}

Hence, the best estimate for $q_0$ is $q"$.

Its accuracy is driven by the \abmcc\ method in finding $\hat E_0=(\varepsilon_x,\varepsilon_y)_{(0,0)}$ from Eq.\,\ref{eq:OTsol}, we note it $\textsf{Cov}(E_0)$, and by the accuracy of the rotation in Eq.\,\ref{eq:q_quote}, we note it $\sigma_\omega$. The estimates $\hat E_0$ and $\hat E$ are strongly correlated, and likely close to each other, as they use the same set of measurements in the same way. Hence, a good approximation for its covariance is the one for the beacon itself, as given by Eq.\,\ref{eq:OTsol}:

\begin{equation}\label{eq:Cov(E0,E)}
\textsf{Cov}(E_0,E)\simeq \textsf{Cov}(E_0) \simeq \textsf{Cov}(E).
\end{equation}

For $\sigma_\omega$, an upper-value can be found and considered uncorrelated with the two others. We assume that the \ADCS' accuracy applies in the whole field of view. The alignment error applies proportionally to the off-axis distance of the beacon, with a maximum at the edge of the field of view. Then, the alignment is improved by the number of stars used in the least-squares method. As a result, the upper-limit can be expressed as:

\begin{equation}\label{eq:sigmaOmega}
\sigma_\omega \simeq \frac{\sigma_\textsf{AKE}}{\sqrt{n^\star}} \frac{\Vert \hat E \Vert}{\alpha_\textsf{FoV}},
\end{equation}

where $\alpha_\textsf{FoV}$ is the angular \emph{half-size} of the field of view and $\Vert \hat E \Vert$ is the angular off-axis distance of the beacon as estimated by the \abmcc.

\subsubsection{Reverse gnomonic projection~}~
Now, the characterization of the field of view is defined by $\textbf{p}_g"$. The \emph{expected} values for $S_{0i}$ are recomputed on this basis: $(x_{0i},y_{0i})_{q"}=\textbf{p}_g"(\vec s_i)$. The \abmcc\ method is processed and yields a solution $(\varepsilon_x,\varepsilon_y)$ given by Eq.\,\ref{eq:OTsol}. Then, the absolute direction of the beacon $\vec b$ and an upper-limit of its error are estimated as follows:

\begin{equation}\label{eq:finalSol}
\begin{aligned}
(x_0,y_0)_{q''} =&~ (x,y) - \hat E, \\
\vec b =&~ (\textbf{p}_g")^{-1}(x_0,y_0), \\
\textsf{Cov}\left(\vec b \right) \simeq &~ \sigma_\omega^2\,I_2 + 4\textsf{Cov}(E).
\end{aligned}
\end{equation}

\emph{Note:} the coefficient 4 before $\textsf{Cov}(E)$ is obtained from $\textsf{Cov}(E_0+E)=\textsf{Cov}(E_0)+\textsf{Cov}(E)+2\textsf{Cov}(E_0,E)$, considering the approximation in Eq.\,\ref{eq:Cov(E0,E)}. 

\subsection{Discussion on the theoretical approach}

One can note that the expression in Eq.\,\ref{eq:OTsigma} tends to zero. But it only relates to the geometrical constraints for the intersection of multiple circles (angular distances) to one single beacon image. Indeed, it expresses the auto-calibration of the field of view due to the presence of stars. It is pretty close to the techniques used in ground astrometric reduction of space images, except that we looked for a formulation that minimises computations: the underlying physical limitations of the optical system still apply, as it is expressed in the covariance upper-limit $\textsf{Cov}(\vec b)$ in Eq.\,\ref{eq:finalSol}, that also benefits from the use of multiple stars in the field of view. The true limit of the method will be estimated statistically for this experiment. For any other mission, and depending on the mission planning, it could also be calibrated using star clusters, as it has already been done for a spacecraft like Cassini, and for sure many others.

The weighting matrix $W$, inverse of $V=\textsf{Diag}\{\textsf{Var}(\nu_i)\}$ could also be improved by using the set of \snr s $\{\sigma_\textsf{SNR,i}\}$. This, along with alternative centerfinding techniques, shall be explored in an advanced version of the experiment on OPS-SAT. Yet, in the image pre-processing, any particular star $k$ with $\{\sigma_\textsf{SNR,k}\}<3$ will be ignored and the linear system in Eq.\,\ref{eq:A=CE+N} will be built with the remaining stars only. Of course, the \snr\ for the beacon itself, even weak, could not be ignored. However, the detailed specification using $\{\sigma_\textsf{SNR,i}\}$ is not ready yet and will not contribute to the commissioning results.

\section{OPS-SAT Experiment ID.156}\label{ID156}

Here, we identify the properties of ESA's OPS-SAT CubeSat that are related to our astrometry experiment. The specific challenges of our experiment with OPS-SAT are presented to justify the progressive deployment. Then, the mission preparation, the implementation as a wrap-up sequence of the \abmcc's equations embedded in OPS-SAT and, eventually, the post-processing are detailed.

\subsection{ESA's OPS-SAT mission}

\begin{figure}[t]\centering
   \includegraphics[width=0.95\linewidth]{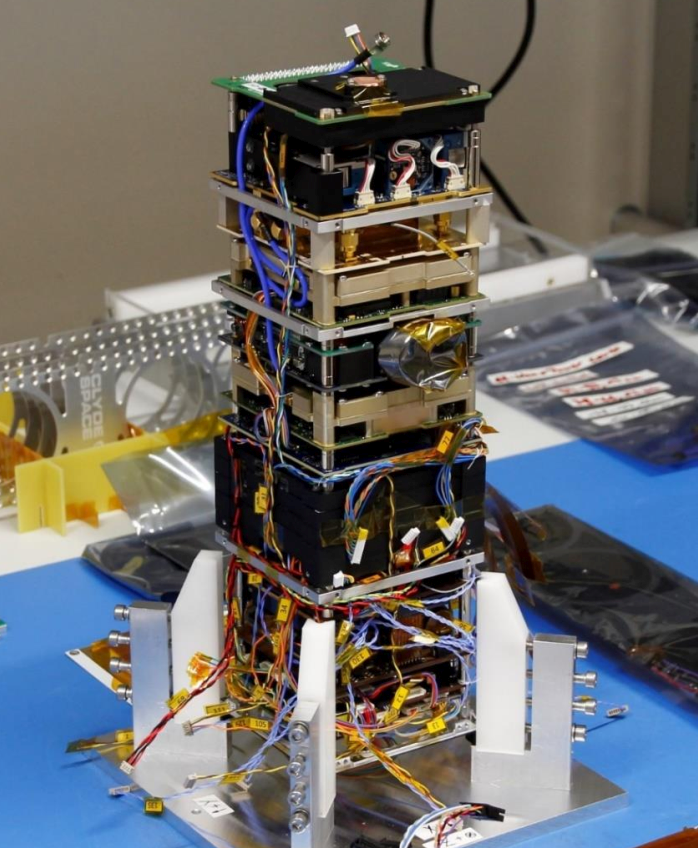}
   \caption{\label{fig:opssat1}ESA's OPS-SAT launched on $18^\text{th}$ Dec 2019. Credit ESA.}
\end{figure}

OPS-SAT (Fig.\ref{fig:opssat1}), also called by ESA ``the flying software lab", is a 3-Unit CubeSat (main body within 30\,cm$\times$10\,cm$\times$10\,cm) and the first nanosatellite directly owned by ESA and controlled by ESA/ESOC in Darmstadt, Germany \cite{Evans2016}. It flies in low Earth orbit, on a polar 6-18 Sun-synchronous orbit. It is equiped with a full set of sensors and actuators, in particular the imager IMS-100 and the fully integrated \ADCS\ iADCS-100, both by Berlin Space Technologies GmbH (BST). The bus is based on NanoMind on-board computers by \textsc{GOMspace} and a software interface was prepared by GMV Poland. The main solar panels, once deployed, consist of five strings of 30\,cm$\times$10\,cm facing $-X$ axis. Although the experiment aims at saving datalink budget by processing the needed data on board, it will take full advantage of OPS-SAT's high datalink in S-band and X-band to download and check on ground the results of the on-board experiment.

The imager is aligned with the satellite's longitudinal $-Z$ axis. The \ADCS\ uses a star tracker, whose axis is in the satellite's $(+X,+Y)$ plane. According to private communications with a star tracker manufacturer for CubeSats, the \AKE\ can be expected within 15\,arcsec transversely to its line of sight, but it likely increases by a factor of 5 to 6 for the axis rotation error. For OPS-SAT, whose imager is not aligned with the star tracker, the start tracker's \AKE\ to be considered in the field of view is likely not isotrope and could reach up to 90\,arcsec in some directions of the image (not considering estimation filters that are embedded in the \ADCS).

The IMS-100 was developed essentially to image the Earth surface. Hence, the calibration and acceptance tests for this device addressed rather the risk of image distortion and not explicitly the sensitivity. The raw images are coded in the IMS-100 on 12 bits per pixel, using an RGGB Bayer pattern, on a 2048$\times$1944 matrix of pixels. They are converted on board in 3-color 8 bit PNG format, by default, but the raw images can still be accessed by an experiment. The behavior at imaging stars is still unknown.

OPS-SAT allows to fly algorithms directly written for Linux shell, in JAVA, in Python or, in the case of this paper, in C++. To this aim, the Institute of Communication Networks and Satellite Communications, from Graz University of Technology, Austria, prepared the \sepp\ module as a library of high-level functions to interface with various OPS-SAT systems, like the imager and the \ADCS. Hence, the experimenter's code uses functions of \sepp\ only and not any functions of the sensors' and actuators' interfaces directly. \sepp\ is of great help for the experimenter who, nevertheless, still needs to understand and care of the correct chronology in the activation of OPS-SAT's systems and of their detailed status before proceeding from one step to the next. A full framework in Eclipse IDE was prepared to cross-compile the experiment code from the experimenter's environment (Intel, AMD) to the target environment, namely OPS-SAT with an ARM processor.

An ``experimenter'' can be any kind of team around the world, whose idea is deemed feasible and relevant with OPS-SAT operations. The experiments are pieces of software to be run on board OPS-SAT, under the principle of ``develop, fly, improve'' cycles. When the experiment is ready, a software review is performed. Then, the orbital deployment is decided (uploading the software and planning its execution) and a validation report terminates the cycle, to assess if debugging or a new orbital deployment is required.

OPS-SAT is fully operational since mid-2020 and a second CubeSat, OPS-SAT 2, is under study.

\subsection{Experiment Id.156}

We submitted the idea to test the \abmcc\ on OPS-SAT and it was approved for implementation on 21$^\text{st}$ Dec 2020. Our experiment was assigned the identification number ``156''. The idea is to characterize the performance of the \abmcc\ with star-fields, where the central star plays the role of the beacon. Considering the existing feed-back on OPS-SAT's stability with the iADCS-100 and, reversely, the lack of experience at imaging star-fields with the IMS-100, we have anticipated successive versions for the on-board code and developed also a tool for the mission preparation.

Regarding the inertial pointing, the OPS-SAT team provides recorded seminars, prepared by TU Graz, Austria, to teach the interface with the iADCS via \sepp. Results of ``high-level'' pointing modes controlled during the commissioning phase are reported. For instance, the ``Target Pointing Nadir'' mode, i.e. -Z axis to the surface, meaning that the star tracker points tangentially to the surface, illustrates the needed execution times: 200 to 1200\,s for attitude acquisition, then less than 300\,s for the achievement of the tracking. However, with the ``Inertial Pointing'' mode, some tests failed at acquiring the attitude even after 40\,mn, while other tests took only 50\,s, illustrating the dispersion of the performance. The reasons are not provided but, as a star tracker cannot recognize 100\,\% of the sky and may suffer from stray-light, one could suspect that the issue is related to particular pointed directions: because the imager and the star tracker point to perpendicular directions, the latter may suffer from stray-light by the Sun, the Earth's albedo or even the Moon, or from occultation by the Earth. An important concern of our experiment will be to care of the elongation angles of the imager's and star tracker's line of sights to the Sun, the Earth and the Moon.

Although IMS-100 is fit to Earth observation, and considering that the same technology is used in the iADCS-100 star tracker, one can expect sufficient \snr s for stars up to the 5$^\text{th}$ magnitude with short exposure. According to preparatory discussions with OPS-SAT, with exposure times from 100\,ms and longer, dim stars should be visible as well, possibly smeared, and with exposure times of 200\,ms, 5 images within 1\,s could possibly be stacked by a built-in function. Then, however, the platform pointing is also reported to degrade progressively, i.e. to drift away from the target attitude. In addition, the \psf\ of a star will be mosaiced by the Bayer pattern of the CCD. But the \psf\ will expand over $\lambda/D\sim$ 7 to 10\,arcsec (with $D\sim$12\,mm lens), much smaller than the RGGB pattern (30\,arcsec$\times30$\,arcsec). As a result, a \psf -matching centerfinding technique could become unefficient with stars dimmer in green than in blue or red. We will then take care to download subwindows of the stars in their raw format, not demosaiced, for investigation on ground.

Nevertheless, CubeSat hardware is not expected to offer the same stability and sensitivity like the hardware for larger space missions. Such limitation is compensated by a flexibility in planing and operating experiments on OPS-SAT, that was never offered in the past. In order to adapt, our experiment is designed in a progressive complexity:
\begin{itemize}
\item Series version 0.x: characterize the stability and the sensitivity of OPS-SAT to image star-fields up to magnitude 7.1, then assess the feasibility of the \abmcc\ with this hardware;
\item Series version 1.x: commission the \abmcc\ algorithm;
\item Series version 2.x: perform intensive campaigns on various star-fields to gather statistically meaningful datasets, including with various stray-light conditions;
\item Series version 3.x: explore alternative centerfinding techniques and weighting methods.
\end{itemize}

The experiment presents some feasibility uncertainties. In our risk assessment, we suggest to re-orient the experiment in case of a feasibility issue: then, we would study the requirements for CubeSat components, and design in-flight acceptance tests, about the needed performances of the \ADCS\ and the imagers for astrometry.

\subsection{Mission preparation}

\begin{figure*}[t]\centering
   \includegraphics[width=0.85\linewidth]{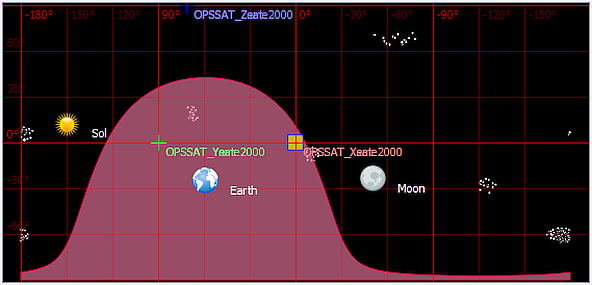}
   \caption{\label{fig:star-fields}Sensor View (within CNES' VTS tool), when aligned with ICRF frame, displaying the selected star-fields and, at a particular epoch time, the area occulted by the Earth (pink) and the directions of the Sun and the Moon.}
\end{figure*}

First, we selected 6 star-fields in the sky. Then, our mission preparation tool computes the needed elongation angles all over the OPS-SAT future orbits. The results are displayed in VTS \cite{vts}, a specialized free tool for space data published by the French space agency CNES.

\subsubsection{Star-fields~}~

\begin{table}
\renewcommand{\arraystretch}{1.3}
\caption{\bf Main properties of the six star-fields}
\label{tbl:star-fields}
\centering
\begin{tabular}{|l|c|c|c|c|}
\hline
\bfseries \# & \bfseries central & \bfseries ICRF & \bfseries Nb.of & \bfseries $m_v$\\
 & \bfseries star & \bfseries direction $(^\circ)$ & \bfseries stars & \bfseries range\\
\hline\hline
1 & HIP  20889 & ( 67.15,  19.18) & 18 & 3.5\dots 7.1\\
2 & HIP  57380 & (176.46,   6.53) & 20 & 2.8\dots 7.1\\
3 & HIP  60260 & (185.34, -60.40) & 84 & 0.0\dots 7.1\\
4 & HIP  76470 & (234.26, -28.14) & 15 & 4.0\dots 7.0\\
5 & HIP  96662 & (294.79,  68.65) & 18 & 4.6\dots 7.1\\
6 & HIP 114939 & (349.21,  -7.73) & 30 & 2.9\dots 7.1\\
\hline
\end{tabular}
\end{table}

In order to accommodate the risk of stray-light all over the year, we identified six star-fields: four close to the ecliptic plane and 2 close to the poles (see Fig.\,\ref{fig:star-fields}). The star-fields are named after the central star and are listed in Table \ref{tbl:star-fields} with their main properties. The full catalog of stars considered in the experiment is listed in appendix \ref{ch:catalog}. Each star in the catalog is within an angular radius of 5$^\circ$ from one central star, it is brighter than magnitude 7.1 in blue, green or red and its type and immediate vicinity are deemed unambiguous with regard to a centerfinding technique. The stars were selected through the use of the SIMBAD database search engine \cite{cds}.

\subsubsection{Elongation angles~}~

\begin{figure*}[t]\centering
   \includegraphics[width=\linewidth]{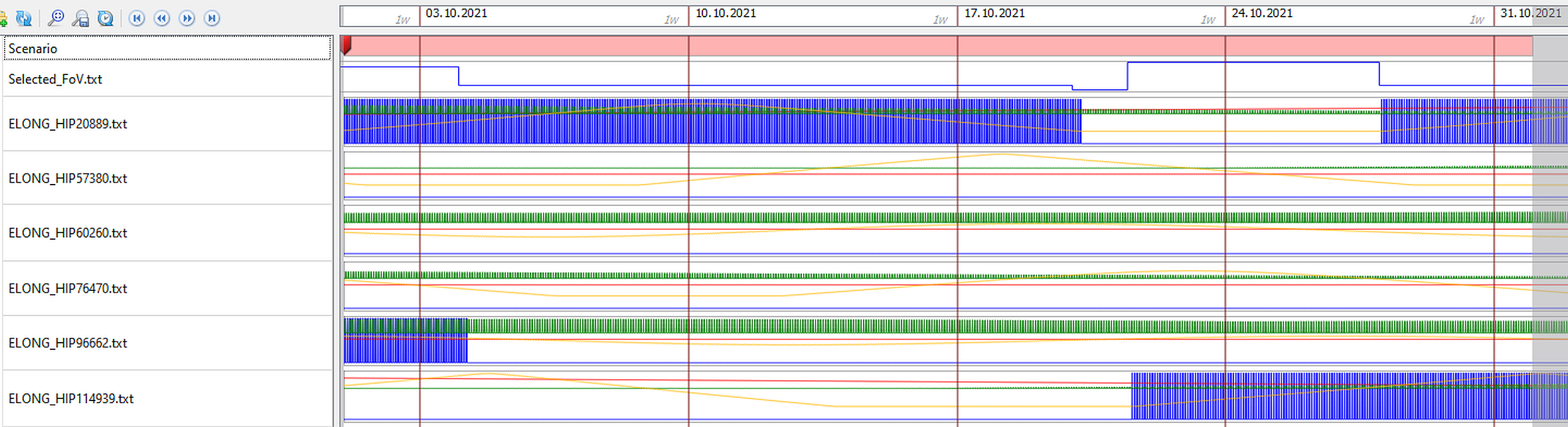}
   \caption{\label{fig:elongations}Elongations over one full month to Sun, Earth, Moon  for each star-field, with its observability flag (blue blocks). Top blue square curve: manually selected star-field.}
\end{figure*}

We compute the angles from the Sun, the Earth-center and the Moon to the direction of each central star of the star-fields. To this aim, the ephemerides of OPS-SAT are built from the propagation of the latest \tle\ available. Although \tle\ are valid for a few days strictly speaking, the TLE-based propagator embedded in VTS includes the Earth J2 effect at the accuracy we need to plan the operations, i.e. the orbit remains Sun-synchronized on 6-18 for a full year ahead.
Then, these three angles are compared to the thresholds that define the cone-avoidance angles for the imager. The exact values are still in discussion. In the presented plots, we consider 95$^\circ$ from the Sun, 50$^\circ$ from the Moon and 50$^\circ$ from the limb of the Earth (not considering day or night side, at this stage). When all three angles are beyond the cone-avoidance angles, an ``observability'' flag is switched on for the star. The Fig.\ref{fig:elongations}, presents the resulting plots for each star during one month, with the observability being the square curve (blue).

\subsubsection{Manual decision~}~

From the shown blocks of observability, we decide manually the star-field that should be observed during each period of time. It is labelled ``Selected FoV'' in Fig.\,\ref{fig:elongations} and shown on the top. At least one star-field should be observable at any time of the year, otherwise we would search for additional star-fields if it becomes incompatible with OPS-SAT planning. From this selection, the quaternion to be sent to the \ADCS\ is computed. At the moment, they are fixed quaternions for each star-field with the aim of aligning one imager axis with the local ICRF longitude to ease the human recognition of the star-field in the pictures. However, the quaternions may be computed to fit with the best rotation about the line-of-sight of the imager to accommodate the attitude tracking by the star tracker. At the time of writing, this additional optimization is still to be discussed with OPS-SAT team.

\subsubsection{Observation planning~}~

\begin{figure}[t]\centering
   \includegraphics[width=0.95\linewidth]{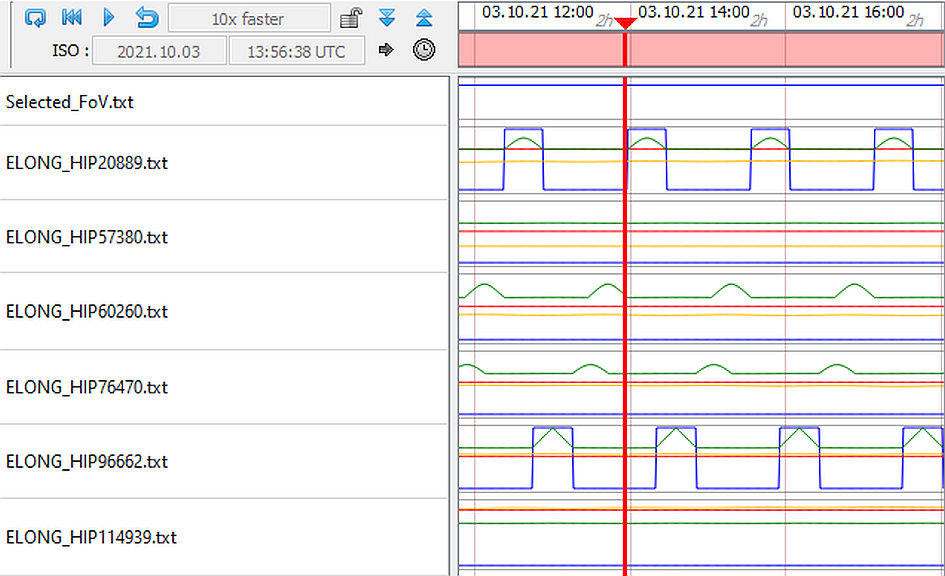}
   \caption{\label{fig:planning}Detailed observability for each star-field. An observability slot typically lasts about 30\,mn.}
\end{figure}

Then, the exact possible periods of time for observation are computed. Indeed, the blue blocks only tell the period of observability. But, when zooming in at a particular epoch time (Fig.\,\ref{fig:planning}), the detailed elongations to the Earth limb over the OPS-SAT orbits, make the observability alternate. Hence, the detailed planning of possible executions of the experiment are output to OPS-SAT team for orbital deployment and planning, as well as stored on-board as an input for the experiment itself.

\subsection{Initial implementation}

The code of our experiment is written in C++. This language was preferred to Python or Java that are also suggested by OPS-SAT, but deemed less portable to various non-CubeSat platforms. In order to test our code on ground, without the actual OPS-SAT hardware, the \sepp\ module and the IMS-100 had to be shunted: their API ``.h'' headers are kept but a new notional function is coded for every function called by our experiment (16 for \sepp\ and 5 for the imager). Whenever the new notional function is called, test data are returned and an entry is added in a test log file, that confirms our variables circulate as expected. When the code is deemed satisfactory, the shunts are simply removed and the same code is cross-compiled in OPS-SAT Eclipse IDE for final check before delivery.

The code itself is currently organized in a sequence of four sets of operations:

\begin{enumerate}
\item Pointing:
\item[ ]

The iADCS-100 is initialized by the experiment itself on the basis of the on-board clock, updated from the ground by the OPS-SAT operation team and provided by the Linux operating system. Then, the high-level ``inertial pointing'' mode is requested to the next target in the observation planning, in association with its prepared quaternion, as computed in mission preparation and uploaded. A strategy to command the iADCS-100 is still in discussion, due to the risk of failure for the star tracker to catch a known constellation. The current approach is to turn on the star tracker but to pre-align on the basis of the sun sensors and magnetometers only, then monitor the reported attitude, then watch whether the iADCS-100's quaternion has reached the target and, in addition, if the star tracker also computes a quaternion by itself (as it can be accessed directly if acquired). If so, the reported attitude will be expected to be in line with the \AKE. Then, a watchdog will limit this process in case the star tracker's quaternion is never acquired.

\item[ ]
\item Imaging:
\item[ ]

During the IMS-100 characterization stage for the \abmcc, a series of picture with various sets of gains and exposure times will be performed, stored and downloaded for ground analysis. The experiment does not attempt to go further at this stage. The \ADCS -reported quaternions are still monitored just before and just after commanding each image. The result of the characterization stage may yield to stack a series of 5 pictures exposed 200\,ms each.

\item[ ]
For the commissioning phase of the \abmcc, series of two pictures will be taken (we do not consider the possible decision of stacking multiple images, yet). If one of the picture contains a cosmic ray that disturbs the centerfinding pre-processing, the other will be used instead. Otherwise, both pictures will be used. Whether a cosmic ray is detected is part of the pre-processing, presented below. The full pictures will be downloaded for analysis (example of an expected star-field is shown in Fig.\,\ref{fig:fov}).

\item[ ]
At the next stages of the experiment, the main program will loop, once the pointing is established: take a new picture as long as the experiment is still running and the \ADCS -reported quaternion is still in range with imaging the central star, and process the image. We are interested in the processing results, even if the satellite drifts away from the target quaternion, to assess the off-axis effects on the beacon measurements. At these stages, the full images are no longer downloaded (but the subwindows are).

\begin{figure}[t]\centering
   \includegraphics[width=0.95\linewidth]{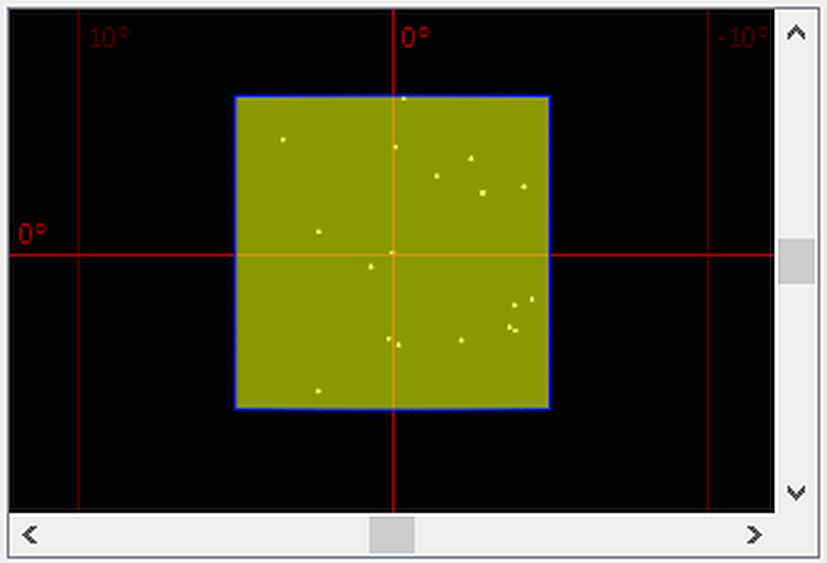}
   \caption{\label{fig:fov}Star-field HIP96662 as expected in the imager field of view (computed in VTS SensorView). The square is $10^\circ\times10^\circ$.}
\end{figure}

\item[ ]
\item Pre-processing:
\item[ ]

Considering the Bayer pattern of the imager, the experiment derives the value for each color in each pixel by a demosaicing method and sums up all three channels to form a grey raw image of the same original size.

\item[ ]
At the timestamp of the image, an interpolated quaternion $q_\textsf{ADCS}$ is computed, as reported by the \ADCS. The gnomonic projection $\textbf{p}_g$, based on $q$ from Eq.\,\ref{eq:q_initial} is processed for all cataloged stars of the star-fields (including the beacon, central star), resulting in a set of \emph{expected locations} $\{(x_{0i}, y_{0i})\}$. The associated subwindows $(x_{0i}\pm3\sigma_\textsf{AKE}, y_{0i}\pm3\sigma_\textsf{AKE})$ are extracted and stored for later download (in raw non-demosaiced format). The \snr\ for each subwindow is computed, considering that the level of ``noise'' is given by the median of the pixels and the level of signal by the brightest pixel. If the SNR is lower than 2, the subwindow is discarded. If the SNR of a subwindow is highly greater than the one of a previous picture (at least 2 pictures are taken), this subwindow is suspected to have got hitten by a cosmic ray and is discarded too.

\item[ ]
Then, the centerfinding method is run on each subwindow, which results in a set of \emph{measured locations} $(x,y)$ and $\{(x_i, y_i)\}$. At the commissioning stage, we will only consider a photometric barycenter of the subwindow after removal of the median value. If the method proves to be sufficient to retrieve the beacon with an accuracy better than the size of one pixel, we will proceed with the same centerfinding technique. Otherwise, we will first implement a better centerfinding technique (\psf -fit, pattern-based correlation). Such better techniques can be explored anyway on ground, based on the already acquired dataset (that includes the raw subwindows) or on board in the last stage of the experiment.

\item[ ]
\item \abmcc:
\item[ ]
The \abmcc\ technique is, then, processed: the gnomonic projection is consolidated in two steps. First, using \abmcc\ to consolidate the line of sight and resulting in the gnomonic projection $\textbf{p}_g'$ based on the quaternions $q_1$ and $q'$ from Eq.\,\ref{eq:q_quote}. Second, the cataloged stars are re-projected with $\textbf{p}_g'$ into the set of \emph{expected locations} $\{(x_{0i}, y_{0i})'\}$, that is fit to the same set of \emph{measured locations} $\{(x_i, y_i)\}$ like previously, yielding a rotation angle of the field of view $\omega$ given by Eq.\,\ref{eq:omega} and an upgraded gnomonic projection $\textbf{p}_g''$ based on the quaternions $q_2$ and $q''$ from Eq.\,\ref{eq:q_dblquote}.

\item[ ]
Then, the cataloged stars are, one last time, re-projected in the field of view with $\textbf{p}_g''$, yielding an upgraded set of \emph{expected locations} $\{(x_{0i}, y_{0i})''\}$. The \abmcc\ is processed again with this new projection, but using the same set of \emph{measured locations} $(x,y)$ and $\{(x_i, y_i)\}$, which results in the estimate of $(x_0,y_0)=(x,y)-\hat E$ for the beacon in the field of view (here the central star of the star-field), with its covariance $\textsf{Cov}(E)$.

\item[ ]
Finally, the image of the beacon is converted in a direction on the sky with its covariance, by the reverse gnomonic projection $(\textbf{p}_g'')^{-1}$ expressed in Eq.\,\ref{eq:finalSol}.

\end{enumerate}

\subsection{Ground processing}

Our OPS-SAT experiment Id.156 is still a work in progress at the time of writing. Nevertheless, as far as possible, it is anticipated to produce the following post-processing:
\begin{itemize}
\item Chronograms of ancillary data, for every experiment run, over run's duration:
	\begin{itemize}
	 \item \ADCS -reported quaternions (sampling at 0.1\,Hz during slew, before and after each image during imaging);
	 \item Statistical evolution compared to inertial pointing commands;
	 \item Corresponding elongations to the Sun, the Earth and the Moon for the imager and the star tracker.
	 \end{itemize}

\item[ ]
\item Before \abmcc\ commissioning, for \ADCS\ and imager characterizations:
	\begin{itemize}
	\item Full-frame images: noise vs. exposure time correlation, noise removal, \psf\ vs. RGGB Bayer pattern assessment, stability and sensitivity initial assessments;
	\item Subwindows: \ACE, \AKE\ and \AKE's isotropic property assessments (before calibration of the field of view with the \abmcc).
	\end{itemize}

\item[ ]
\item Systematic ground processing for every single on-board image:
	\begin{itemize}
	\item Subwindows: automatic monitoring of individual star centerfinding and \snr, and manual investigation in case of strong deviation;
	\item Field of view \abmcc -calibration: monitoring of $\hat E_0$, $\textsf{Cov}(E_0)$, $\omega$ from Eq.\,\ref{eq:q_quote} to \ref{eq:Cov(E0,E)};
	\item Beacon \abmcc -estimates: $(x_0,y_0)$ and \textsf{Cov}(E) from Eq.\,\ref{eq:OTsol} (in field of view), $\vec b$ and $\textsf{Cov}(\vec b)$ from Eq.\,\ref{eq:finalSol} (on the sky);
	\item In-image Centerfinding performance assessment;
	\item In addition, for \abmcc\ commissioning: ground re-processing of the subwindows from the downloaded full-frame images; manual assessment of quality (cosmic, hot pixels, \snr, ambiguities among stars), individual star centerfinding assessment from the subwindows.
	\end{itemize}

\item[ ]
\item Overall analysis of the datasets:
	\begin{itemize}
	\item Statistical properties of \textsf{Cov}(E) and search for correlations with $\sigma_\textsf{out}$ expressed by Eq.\,\ref{eq:OTsigma};
	\item Search for correlation with straylight conditions;
	\item Radial optical aberration assessment based on \abmcc -calibrated star-field.
	\end{itemize}

\end{itemize}

An original part of our experiment will be to link the experiment mission profile, its mission preparation and its post-processing altogether. In this aim, the on-board code will directly format log-files for VTS visualization in our mission preparation tool. Such VTS-formatted log-files will target the following data:
\begin{itemize}
\item \ADCS\ requested and reported quaternions;
\item \abmcc -computed beacons in the field of view and on the sky;
\item \abmcc -calibration of the field of view;
\item Number of stars taken into account (acceptable \snr).
\end{itemize}

The corresponding datavolume for these log-files is still small, considering the OPS-SAT datalink. The effect will be an immediate quick analysis of the results, hence more flexibility in the ``develop, fly, improve" cycles offered by OPS-SAT.

\section{Potential applications}\label{applications}

Although the \abmcc\ is meant for astrometry, it can also be used as an astrometric tool to characterize \ADCS\ and imagers. Here are examples in both application contexts.

\subsection{Astrometry to commission \ADCS\ and Imager}

The \ADCS\ performance strongly depends on the rest of the platform, its preferred orientation (OPS-SAT observes mainly with $-Z$ axis to the Earth, while the \ADCS' star tracker points perpendicularly to this direction) and the reliability of space-environment models. It makes it very difficult to test for acceptance. Interestingly, our first 3-Unit CubeSat at Paris Observatory, PICSAT, launched in 2018 (Nowak 2018 \cite{Nowak2018}) used the same \ADCS\ like OPS-SAT (BST iADCS-100), but provided by Hyperlink Technologies B.V., Netherlands, with the star tracker ST200 aligned with the line of sight of the telescope ($+Z$ axis). In 8 weeks of operations, the \ADCS\ was never able to align the satellite with the star $\beta-$Pictoris, its primary target, despite the detumbling reached less than $0.1\,^\circ/s$, well below the manufacturer's request. As already reported, OPS-SAT has also experienced delays, and even failures, in establishing an inertial pointing, despite a strong endeavour in the development of high-level pointing modes, that could result from the same root causes.

These two examples remind the difficulty to test the \ADCS\ on ground, and to commission it during flight. In particular, monitoring the quaternions is not enough if an absolute reference is not available, which can be provided by an imager after careful positioning of each image against the sky.

The rigorous acceptance is also a challenge for a CubeSat imager whose market-oriented design is best at observing the Earth surface in daylight, but maybe not characterized for other uses (although it could fit to such uses). As a highly-integrated subsystem, presented as a ``COTS'' component (commercial-off-the-shelves), it comes sometimes with a poor datasheet, few test results and poor customer support. Hence, the elementary and integration tests at CubeSat level require a good preparation and specialized tooling, with still a risk of delay. It also needs to anticipate for the search of a good setup and a software-based geometric re-calibration and picture cleaning in flight.

As soon as both an \ADCS\ and an imager are present in a satellite, our \abmcc\ method can serve as a guidance for acceptance tests on ground and, later, in flight.

\subsection{Autonomous astrometry in CubeSat}

\subsubsection{On-board orbit determination~}~
We have developed a solution for autonomous orbit determination for CubeSat in deep space cruise (Segret 2019 \cite{Segret2019b,Segret2019a}, 2021 \cite{SegretarXiv}). Several teams also work on the development of autonomous navigation solutions, especially because of the likely rise of deep space CubeSats, as stand alone or accompanying probes. Optical measurements of distant beacons are limited by the accuracy of these beacons that is proportionally translated into the overall accuracy of the orbit determination. The optical accuracy is usually assumed to be the size of a pixel, which is far from true as we have seen in this paper, due essentially to \AKE, unless a specific image processing. All these attempts are immediate possible applications of the presented  \abmcc\ experiment, with the expectation of a sub-pixel and even a sub-arcsec accuracy.

\subsubsection{Close-to-Sun Asteroids~}~
Another application for astrometry is the detection of celestial bodies on Sun-side as seen from the CubeSat, i.e. below $90\,^\circ$. CubeSats are good candidates for such missions in Earth orbit or at some distance (actually up to the Moon, provided a datarelay is available). Indeed, this goal needs space observation and CubeSats represent a low cost, high risk paradigm. Obviously, stray-light is here the major concern and imagers like the one on OPS-SAT would not fit to this use. Nevertheless, a different set of sensor and baffle can be considered and the \abmcc\ will be useful as a pre-processing technique in the domain of Space Situation Awareness (SSA) for early detection, assessment and warning system.

\subsubsection{Debris reconnaissance~}~
In the same aim, debris in Earth orbit may be difficult to locate and a broad endeavor is ongoing to propose debris removal solutions. It can be interesting to develop a triangulation method between a space tug and an accompanying nanosatellite (or several) to approach the debris autonomously from far range. Then, an on-board processing of each platform's images is highly desirable and the \abmcc\ can provide the required level of accuracy.

\section{Summary}

This paper has detailed the \abmcc\ method, an on-board processing to derive the absolute direction of a beacon in an image, along with an upper-limit for its accuracy. The theoretical expression of this upper-limit makes the method very promising for applications at CubeSat scale, whose wide-angle camera can easily image about 10 bright stars along with the beacon, despite a weak attitude knowledge. The implementation of the \abmcc\ on board an actual ESA CubeSat has been presented and will make possible, from 2022, to quantify the performance of the method, its sensitivity and, as a bonus, the stability and accuracy of the CubeSat pointing itself.

Applications to on-board orbit determination are first targeted. In addition, the in-flight testing of the \abmcc\ method has been an opportunity to define a commissioning process for a CubeSat \ADCS\ and for an imager used as a NavCam. The ground testing being such a challenge with standardized devices of the New Space, CubeSat integrators may feel quiet powerless when dealing with their newly delivered hardware. Thus, we strongly encourage them to share their experience at using the \abmcc\ method with their own hardware, on ground or in flight, with the hope that it will improve the dialog with the device manufacturers.

\appendices{} 


\section{Mathematical developments}
\subsection{Development of \textsf{Cov}(V) for Eq.\,\ref{eq:OTopt}}
Considering the centerfinding process to be dominated by an uniform, zero-mean, Gaussion error $\sigma_\textsf{in}$ on both dimensions:
\[\begin{aligned}
\forall i & \in\{1..n^\star\}, \textsf{Var}(\nu_i) \\
 &= \textsf{Var}(a_i^2-a_{i0}^2), \text{ where } a_{i0} \text{ not a random variable}\\
 &= \textsf{Var}(a_i^2) \\
 &= \textsf{Var}(x^2 -2xx_i +x_i^2 + y^2 -2yy_i +y_i^2).
\end{aligned}\]
The random variables $x$ and $x_i$ (resp. $y$ and $y_i$) being independent and not correlated, $\textsf{Cov}(x,x_i)=0$ (resp. $\textsf{Cov}(y,y_i)=0$). Hence,
\[\begin{aligned}
\textsf{Var}(\nu_i) &= \textsf{Var}^2(x) +\textsf{Var}^2(x_i) + \textsf{Var}^2(y) +\textsf{Var}^2(y_i) \\
 &= 4(\sigma_\textsf{in}^2)^2,
\end{aligned}\]
because the variances of $x$, $y$, $x_i$ and $y_i$ are taken equal in the context of Eq.\,\ref{eq:OTopt}.

\subsection{Development of \textsf{Cov}(E) for Eq.\,\ref{eq:OTsigma}}
Special case, where $B(x_0,y_0)$ is at the barycenter of a cloud of stars whose coordinates $\{(x_{0i},y_{0i}), i=1\dots n^\star\}$ have a standard deviation $d$ in both dimensions.

\[W=\left(\frac{1}{4\sigma_\textsf{in}^4}\right) \, I_{n^\star},\]

\[C = \begin{pmatrix}
\vdots & \vdots \\ -2(x-x_{0i}) & -2(y-y_{0i}) \\ \vdots & \vdots
\end{pmatrix}_{i=1\dots n^\star},\]

\[C^T W C = \begin{pmatrix} C_1 & C_0 \\ C_0 & C_2 \end{pmatrix},\]

where:
\[\begin{aligned}
C_1 &= \sum_{i=1}^{n^\star}-2(x-x_{0i})\left(
\frac{-2(x-x_{0i})}{4\sigma_\textsf{in}^4}
\right) \\
    &= \frac{1}{\sigma_\textsf{in}^4}\sum_{i=1}^{n^\star}(x-x_{0i})^2
\end{aligned},\]
and, considering that B is at the barycenter with the standard deviation $d$ for the cloud of stars:
\[C_1 = \frac{n^\star d^2}{\sigma_\textsf{in}^4}.\]
Similarly:
\[\begin{aligned}
C_2 &= \sum_{i=1}^{n^\star}-2(y-y_{0i})\left(
\frac{-2(y-y_{0i})}{4\sigma_\textsf{in}^4}
\right) \\
    &= \frac{1}{\sigma_\textsf{in}^4}\sum_{i=1}^{n^\star}(y-y_{0i})^2 \\
    &= \frac{n^\star d^2}{\sigma_\textsf{in}^4}
\end{aligned}.\]
And the anti-diagonal elements become null:
\[\begin{aligned}
C_0 &= \sum_{i=1}^{n^\star}-2(x-x_{0i})\left(
\frac{-2(y-y_{0i})}{4\sigma_\textsf{in}^4}
\right) \\
    &= \frac{1}{\sigma_\textsf{in}^4}\sum_{i=1}^{n^\star}(x-x_{0i})(y-y_{0i}) \\
    &= 0
\end{aligned}.\]
Hence:
\[\textsf{Cov}(E) = \left(C^T W C\right)^{-1}
 = \frac{\sigma_\textsf{in}^4}{n^\star d^2}\, I_2,\]

then finally the Eq.\,\ref{eq:OTsigma}, noting $\textsf{Cov}(E)=(\sigma_{\textsf{out}}^2)\,I_2$:
\[\sigma_\textsf{out} = \frac{\sigma_{\textsf{in}}}{\sqrt{n^\star}}\left(\frac{\sigma_{\textsf{in}}}{d}\right).\]

\subsection{Development for the cost function $F(\omega)$ in Eq.\,\ref{eq:F(w)}}

\[
F(\omega) = \sum_{i=1}^{n^\star} \Vert \overrightarrow{S_i \, \Omega(S_{0i})} \Vert ^2
\]

\[\begin{aligned}
\Vert \overrightarrow{S_i \, \Omega(S_{0i})} \Vert ^2 = &
    \left( \cos\omega x_{0i} - \sin\omega y_{0i} - x_i\right)^2 \\
  & + \left( \sin\omega x_{0i} + \cos\omega y_{0i} - y_i\right)^2 \\
\text{(then,} & \text{ at second order)} \\
= & \left( (1-\omega^2/2) x_{0i} - \omega y_{0i} - x_i\right)^2 \\
  & + \left( \omega x_{0i} + (1-\omega^2/2) y_{0i} - y_i\right)^2 \\
= & \left( (x_{0i}- x_i) - \omega y_{0i} - (\omega^2/2) x_{0i}\right)^2 \\
  & + \left( (y_{0i}-y_i) + \omega x_{0i} -(\omega^2/2) y_{0i} \right)^2
\end{aligned}
\]

developing further, keeping only the terms up to the 2nd order and grouping as a polynomial of $\omega$:

\[\begin{aligned}
= &~ \omega^2 \left[ y_{0i}^2 - x_{0i}(x_{0i}-x_i) + x_{0i}^2 - y_{0i}(y_{0i}-y_i) \right] \\
  & + \omega \left[ -2y_{0i}(x_{0i}-x_i) + 2x_{0i}(y_{0i}-y_i) \right] \\
  & + (x_{0i}-x_i)^2 + (y_{0i}-y_i)^2 
\end{aligned}
\]

Then, the full expression of the cost function:
\[\begin{aligned}
F(\omega) = &~ \omega^2 \sum_{i=1}^{n^\star} ( x_{0i}x_i + y_{0i}y_i ) \\
 & + 2\omega \sum_{i=1}^{n^\star} ( y_{0i}x_i - x_{0i}y_i ) \\
 & + \sum_{i=1}^{n^\star} \left[ (x_{0i}-x_i)^2 + (y_{0i}-y_i)^2 \right]
\end{aligned}\]

The coefficient of $\omega^2$ being positive, the polynomial has got a minimum for $\omega$ the solution of $F'(\omega)=0$:

\[\begin{aligned}
F'(\omega) = & 2\omega \sum_{i=1}^{n^\star} ( x_{0i}x_i + y_{0i}y_i ) \\
 & + 2 \sum_{i=1}^{n^\star} ( y_{0i}x_i - x_{0i}y_i )
\end{aligned}\]

\[\begin{aligned}
F'(\omega) &= 0 \\
\Leftrightarrow \omega &= \frac{
\sum_{i=1}^{n^\star} (x_{0i} y_i - y_{0i} x_i) }{
\sum_{i=1}^{n^\star} (x_{0i} x_i + y_{0i} y_i) }
\end{aligned}\]

\section{Quaternion algebra}
A few reminders on basic properties and operations in the quaternion algebra is useful, for instance to check for ambiguous syntax in some languages (e.g. Python). The major concern is that the product of quaternions is not commutative and that the conjugate can only be used, instead of the inverse, with a unit quaternion.

\subsection{Possible notations}

These notations are equivalent:
\[\begin{aligned}
q &= (q_0, q_x, q_y, q_z) \text{, but also sometimes} \\
  &= (q_x, q_y, q_z, q_0) \\
  &= q_0 + q_x \vec i + q_y \vec j + q_z \vec z \\
  &= (q_0 + \vec v) \\
  &= \vert q \vert (\cos \theta/2 + sin \theta /2 . \vec u), where \Vert \vec u \Vert=1 \text{, or} \\
  &= \vert q \vert (\cos \theta/2, \vec u) \text{ with} \\
\vert q \vert &= \sqrt{q_0^2 + q_x^2 + q_y^2 + q_z^2} = \sqrt{q_0^2 + \Vert \vec v \Vert^2}
\end{aligned}\]

Meaning: in the form $(\cos \theta/2, \vec u)$, this \emph{unit} quaternion expresses the rotation by $\theta$ about the axis $\vec u$, the direction of the vector telling the direct rotation, i.e. if $\theta>0$.

\subsection{Important properties}

Note that any vector $\vec v$ can be written as a quaternion $(0+\vec v)$ and any scalar $s$ can be written as a quaternion $s+\vec 0$.

Operations:
\begin{itemize}
\item The addition is associative, commutative, noted ``$+$"
\[ (s+\vec v) + (t+\vec w) = (s+t) + (\vec v + \vec w) \]
\item The product is associative, \emph{not} commutative, noted ``$\otimes$" (cross-product noted ``$\times$", scalar - or dot - product noted ``$\cdot$")
\[ (s+\vec v) \otimes (t+\vec w) = [st - \vec v \cdot \vec w] + [s\vec w + t\vec v + \vec v \times \vec w] \]
\item Conjugate:
\[ q=q_0+\vec v \Longrightarrow q^\star = q_0-\vec v\]
\end{itemize}

Then, $(\vec i,\vec j,\vec k)$ being on orthonormal direct frame:
\[\begin{aligned}
\vec i^2 = \vec j^2 = \vec k^2 &= -1 \\
\vec i \otimes \vec j &= \vec i \times \vec j = \vec k \\
\vec j \otimes \vec k &= \vec j \times \vec k = \vec i \\
\vec k \otimes \vec i &= \vec k \times \vec i = \vec j \\
\end{aligned}\]

And:
\[\begin{aligned}
\vert q \vert = 1 &\Longleftrightarrow q^{-1} = q^\star \\
q \equiv \text{ rotation by } \theta &\Longleftrightarrow q^\star \equiv \text{ rotation by } (-\theta)
\end{aligned}\]

Product:
\[\begin{matrix}
q_1, q_2 \text{ two rotations} \\
\Longrightarrow q=q_2 \otimes q_1 \text{ is a rotation by $q_1$ first, then $q_2$}
\end{matrix}
\]

\emph{Warning:} in python, the operator ``*'' in the package \textsf{pyquaternion} may be confusing, as $q_a * q_b$, there, means a rotation composed by the rotations $q_a$ first, then $q_b$.

Transform of a vector, example:

With $\vert q \vert = 1$, the transform of a vector $\vec v$ by the rotation associated with $q$ is a vector $\vec v'$ whose coordinates are (the frame is \emph{not} transformed, here):

\[ (0+\vec v') = q \otimes (0+\vec v) \otimes q^\star \]

If the attitude of a satellite is expressed by $q$, with ICRF as the reference frame, it means that its body-frame was transformed by the rotation of ICRF associated with $q$, meaning that its axes have ICRF-coordinates given by:

\[\begin{aligned}
\vec X = q \otimes (0+\vec X_\textsf{ICRF}) \otimes q^\star \\
\vec Y = q \otimes (0+\vec Y_\textsf{ICRF}) \otimes q^\star \\
\vec Z = q \otimes (0+\vec Z_\textsf{ICRF}) \otimes q^\star \\
\end{aligned}\]

Hence, if a vector $\vec v$ (with coordinates in ICRF) is projected in the body-frame rotated by $q$, it will appear as a vector $\vec v'$ in the body-frame, whose coordinates are obtained by:

\[ (0+\vec v') = q^\star \otimes (0+\vec v) \otimes q. \]

\section{Star Catalog}\label{ch:catalog}

See Tbl.\,\ref{tbl:catalog}.

Note that the Id. of a star is for internal use only. It is either the Hipparcos catalog number (HIP) or an arbitrary negative number when lacking a HIP number.

{
\begin{table}
\renewcommand{\arraystretch}{1.1}
\caption{\bf Stars selected for the on-board catalog.}
\label{tbl:catalog}
\centering
\begin{tabular}{|c|c|c|}
\hline
\bfseries Id. & \bfseries ICRF & \bfseries $m_v$\\
 			  & \bfseries direction $(^\circ)$ & \bfseries (max)\\
\hline\hline
 19960 & ( 64.202871,  16.214387) &   6.6 \\
   -10 & ( 65.396156,  21.188503) &   7.1 \\
 20789 & ( 66.822714,  22.996329) &   5.4 \\
 20493 & ( 65.884861,  20.982045) &   6.0 \\
 20780 & ( 66.770272,  18.207550) &   6.9 \\
 20889 & ( 67.154162,  19.180429) &   3.5 \\
 20205 & ( 64.948349,  15.627643) &   3.6 \\
 20916 & ( 67.248976,  16.159070) &   6.3 \\
 20877 & ( 67.109877,  16.359674) &   5.0 \\
 20215 & ( 64.978572,  16.522591) &   6.9 \\
 20440 & ( 65.684047,  15.056089) &   7.1 \\
 20732 & ( 66.651549,  14.713786) &   4.0 \\
 21094 & ( 67.804890,  23.346060) &   7.1 \\
 21158 & ( 68.032945,  21.632356) &   7.1 \\
 21408 & ( 68.927355,  19.881749) &   6.3 \\
 21257 & ( 68.408696,  16.323338) &   6.6 \\
 21365 & ( 68.759870,  17.201231) &   6.8 \\
 21588 & ( 69.539371,  16.033253) &   5.8 \\
 20995 & ( 67.535824,  15.637839) &   5.4 \\
    -1 & ( 70.550805,  22.942361) &   7.1 \\
    -7 & ( 68.387722,  18.016722) &   7.0 \\
    -8 & ( 68.386812,  18.016822) &   7.0 \\
 56079 & (172.416647,   7.599375) &   6.7 \\
 56618 & (174.142338,   6.106440) &   6.8 \\
 56445 & (173.591453,   3.060165) &   5.5 \\
 56756 & (174.540965,   8.883792) &   6.2 \\
 56779 & (174.615030,   8.134296) &   5.4 \\
 57328 & (176.321001,   8.258115) &   4.8 \\
 57562 & (176.978734,   8.245895) &   5.3 \\
 57339 & (176.350672,   7.030039) &   7.1 \\
 57380 & (176.464832,   6.529373) &   2.8 \\
 57401 & (176.526216,   7.174190) &   6.9 \\
 57107 & (175.606422,   2.362110) &   7.0 \\
 57888 & (178.087086,   9.948064) &   6.9 \\
 58110 & (178.763052,   8.443945) &   5.6 \\
 58329 & (179.437913,   6.974628) &   6.8 \\
 57636 & (177.279630,   5.183376) &   5.8 \\
 58377 & (179.558042,   3.482009) &   6.7 \\
 58510 & (179.987137,   3.655207) &   5.3 \\
 57757 & (177.673826,   1.764720) &   3.1 \\
 58590 & (180.218293,   6.614321) &   4.6 \\
 58809 & (180.935170,   5.557751) &   6.1 \\
 57644 & (177.303030, -57.691561) &   6.8 \\
 58041 & (178.547920, -57.410197) &   6.0 \\
 58059 & (178.611078, -56.969775) &   6.9 \\
 58465 & (179.848897, -57.167986) &   6.8 \\
 58334 & (179.442643, -60.451347) &   6.8 \\
 57439 & (176.628427, -61.178399) &   4.1 \\
 57808 & (177.804115, -61.846152) &   6.6 \\
 57064 & (175.473819, -60.230810) &   6.8 \\
 57004 & (175.293746, -61.141066) &   6.9 \\
 58326 & (179.416698, -62.448748) &   5.4 \\
 58103 & (178.750049, -63.279180) &   5.7 \\
 58469 & (179.856744, -62.830946) &   6.3 \\
 58427 & (179.698579, -64.339559) &   5.6 \\
 57175 & (175.879964, -62.489396) &   4.3 \\
 58756 & (180.754458, -57.260129) &   6.7 \\
\hline
\end{tabular}
\end{table}

\begin{table}
\renewcommand{\arraystretch}{1.1}
\centering
\begin{tabular}{|c|c|c|}
\hline
 59965 & (184.480362, -57.850149) &   6.7 \\
 59747 & (183.786320, -58.748927) &   2.6 \\
 58642 & (180.371635, -57.503777) &   6.1 \\
 59560 & (183.217097, -59.655312) &   7.1 \\
 59266 & (182.348246, -59.769257) &   7.0 \\
 59528 & (183.105660, -60.067804) &   6.8 \\
 58942 & (181.285704, -61.179558) &   6.7 \\
 59200 & (182.103045, -60.847130) &   6.2 \\
 59396 & (182.771276, -61.277424) &   6.1 \\
 58748 & (180.734803, -62.175287) &   6.8 \\
 58758 & (180.756255, -63.312928) &   4.3 \\
 58867 & (181.080093, -63.165694) &   4.6 \\
 58960 & (181.345450, -62.975752) &   6.7 \\
 59502 & (183.042808, -63.454114) &   7.0 \\
 59517 & (183.091583, -62.950774) &   5.6 \\
 59865 & (184.159280, -63.197906) &   7.0 \\
 60128 & (184.973768, -62.854183) &   6.8 \\
 60009 & (184.609317, -64.003087) &   3.9 \\
 60259 & (185.338033, -62.281632) &   6.6 \\
 59072 & (181.720412, -64.613729) &   4.1 \\
 59488 & (183.003157, -64.510403) &   6.6 \\
 59678 & (183.570523, -64.408519) &   6.2 \\
 59111 & (181.816481, -64.681778) &   6.8 \\
 59457 & (182.934551, -64.948788) &   7.0 \\
 61179 & (188.036661, -56.129540) &   6.5 \\
 60308 & (185.489282, -56.374384) &   5.9 \\
 60870 & (187.139976, -56.407817) &   6.1 \\
 60969 & (187.475830, -56.524936) &   5.8 \\
 60379 & (185.705967, -57.676129) &   5.3 \\
    -5 & (187.819664, -57.081330) &   6.5 \\
 60780 & (186.870150, -58.316447) &   6.3 \\
 61154 & (187.970072, -58.290614) &   7.1 \\
 61084 & (187.791498, -57.113213) &  -0.0 \\
 60781 & (186.870326, -58.991766) &   5.5 \\
 61136 & (187.918042, -59.423923) &   5.5 \\
 61019 & (187.581384, -60.989894) &   6.7 \\
 60455 & (185.907035, -61.629127) &   6.4 \\
 60260 & (185.340039, -60.401147) &   3.6 \\
 60845 & (187.070330, -61.765709) &   6.8 \\
 60861 & (187.107019, -61.794841) &   6.2 \\
 61444 & (188.874366, -61.740805) &   6.6 \\
  -256 & (186.628679, -63.122232) &   4.7 \\
 60771 & (186.853441, -63.788985) &   6.0 \\
 62027 & (190.709440, -63.058625) &   5.3 \\
 60905 & (187.228590, -61.870913) &   6.9 \\
 61158 & (187.983802, -63.505839) &   5.9 \\
 60644 & (186.442779, -64.022088) &   7.0 \\
 60570 & (186.233154, -65.211014) &   7.0 \\
 62179 & (191.145816, -57.286894) &   6.6 \\
 62680 & (192.661215, -58.166283) &   7.1 \\
 62428 & (191.913362, -58.297570) &   6.9 \\
 62084 & (190.868006, -58.902741) &   6.4 \\
 62986 & (193.591655, -58.430615) &   6.6 \\
 62434 & (191.930287, -59.688772) &   1.0 \\
 62268 & (191.408542, -60.981320) &   4.7 \\
 62646 & (192.549173, -60.400530) &   6.7 \\
 61966 & (190.485701, -59.685821) &   4.9 \\
 62732 & (192.824901, -60.329787) &   5.3 \\
 63007 & (193.663261, -59.146701) &   4.5 \\
 62651 & (192.558422, -62.641299) &   6.6 \\
 63489 & (195.131362, -60.375367) &   7.1 \\
 62894 & (193.341225, -60.328489) &   5.8 \\
 \hline
\end{tabular}
\end{table}

\begin{table}
\renewcommand{\arraystretch}{1.1}
\centering
\begin{tabular}{|c|c|c|}
\hline
 62953 & (193.499163, -60.335418) &   6.8 \\
 62913 & (193.406747, -60.357062) &   6.8 \\
 57741 & (177.613653, -62.649381) &   5.7 \\
 60851 & (187.080253, -64.340993) &   6.1 \\
 61443 & (188.872681, -61.841528) &   6.2 \\
 62931 & (193.453831, -60.376241) &   5.9 \\
 58921 & (181.238392, -60.968253) &   6.0 \\
 58901 & (181.188581, -59.253253) &   6.1 \\
 59360 & (182.675277, -56.442319) &   7.0 \\
    -2 & (186.649567, -63.099092) &   1.1 \\
    -3 & (186.651842, -63.099522) &   1.4 \\
 76048 & (232.959292, -32.881115) &   6.5 \\
 74675 & (228.914044, -27.596291) &   6.8 \\
 75272 & (230.688367, -29.341935) &   6.5 \\
   -11 & (234.053774, -28.998852) &   6.9 \\
 76259 & (233.655533, -28.046997) &   5.1 \\
 76470 & (234.256042, -28.135081) &   3.6 \\
 76567 & (234.565879, -28.206656) &   6.3 \\
 77235 & (236.553560, -28.061487) &   6.5 \\
   -12 & (233.289642, -24.490417) &   6.9 \\
 76742 & (235.070410, -23.818076) &   4.0 \\
 77394 & (236.985218, -25.986953) &   7.0 \\
 77840 & (238.402995, -25.327141) &   4.5 \\
 77635 & (237.744771, -25.751295) &   4.6 \\
 77984 & (238.875308, -26.265984) &   5.6 \\
   -13 & (233.287292, -24.489144) &   7.0 \\
   -14 & (238.402989, -25.327136) &   4.6 \\
   -15 & (238.402361, -25.327135) &   7.0 \\
 77985 & (238.877221, -31.083616) &   6.2 \\
 75288 & (230.737740, -26.688762) &   6.5 \\
 77900 & (238.625455, -27.338635) &   6.1 \\
 97831 & (298.198683,  64.176045) &   6.8 \\
 99090 & (301.749940,  66.308018) &   7.0 \\
 98583 & (300.368934,  64.820971) &   5.3 \\
 98658 & (300.584238,  64.634419) &   6.3 \\
 97988 & (298.705870,  64.721225) &   6.9 \\
 99291 & (302.335673,  68.599553) &   7.0 \\
 99435 & (302.737129,  68.272214) &   6.8 \\
101349 & (308.104858,  70.532093) &   6.7 \\
 95081 & (290.167056,  65.714530) &   4.5 \\
 94993 & (289.942247,  64.390807) &   6.4 \\
 96662 & (294.792961,  68.652572) &   6.8 \\
 97122 & (296.076739,  69.337061) &   5.9 \\
 94140 & (287.440857,  65.978515) &   6.3 \\
 93517 & (285.679068,  68.325168) &   7.1 \\
 94505 & (288.516316,  67.115559) &   6.8 \\
 95978 & (292.751123,  70.989341) &   6.1 \\
 92137 & (281.691037,  67.771485) &   7.1 \\
 92672 & (283.253770,  71.787228) &   7.0 \\
114536 & (347.998206, -11.933554) &   7.0 \\
115015 & (349.417015, -11.712948) &   6.3 \\
115839 & (352.021646, -11.449756) &   6.4 \\
113583 & (345.083121,  -8.880515) &   6.8 \\
114600 & (348.244228,  -9.570473) &   7.0 \\
114750 & (348.667386, -10.688719) &   6.1 \\
114855 & (348.972897,  -9.087734) &   3.5 \\
115033 & (349.475891,  -9.182519) &   4.3 \\
115495 & (350.938821,  -8.460103) &   6.7 \\
115115 & (349.740319,  -9.610748) &   5.0 \\
115906 & (352.252764,  -9.266104) &   6.2 \\
113674 & (345.348507,  -7.061154) &   6.2 \\
113781 & (345.635654,  -6.574013) &   6.2 \\
\hline
\end{tabular}
\end{table}

\begin{table}
\renewcommand{\arraystretch}{1.1}
\centering
\begin{tabular}{|c|c|c|}
\hline
114054 & (346.468881,  -7.936714) &   6.7 \\
113996 & (346.290778,  -7.693801) &   5.5 \\
114006 & (346.327164,  -7.754690) &   7.1 \\
114445 & (347.665090,  -5.960274) &   6.8 \\
114724 & (348.580665,  -6.049000) &   2.9 \\
114952 & (349.246463,  -7.160803) &   6.7 \\
114939 & (349.212247,  -7.726501) &   5.1 \\
115257 & (350.170361,  -5.907971) &   6.2 \\
116060 & (352.754760,  -6.288080) &   6.4 \\
116428 & (353.883632,  -7.464431) &   6.4 \\
113686 & (345.382122,  -4.711461) &   5.9 \\
113896 & (345.988640,  -4.794862) &   6.7 \\
114822 & (348.892740,  -3.496379) &   5.5 \\
115220 & (350.066040,  -3.919138) &   6.7 \\
115316 & (350.339756,  -4.672761) &   6.8 \\
115892 & (352.194565,  -5.389421) &   7.0 \\
115953 & (352.383672,  -4.532745) &   6.3 \\
   -20 & (350.938854,  -8.460075) &   7.1 \\
\hline
\end{tabular}
\end{table}
}

\acknowledgments

This work was financed by PSL Université Paris as an outcome of the support provided to various nanosatellite projects by CENSUS, the space pole of PSL Université Paris, hosted by Paris Observatory.

The SensorView and the mission preparation plots come from CNES' free software VTS \cite{vts}. This work has made use of the SIMBAD database, operated at CDS, Strasbourg, France.

\bibliographystyle{IEEEtran}

\thebiography
\begin{biographywithpic}
{Boris Segret}{illustrations/boris}
obtained his Ph.D on the subject of autonomous navigation in deep space. He received an M.Sc. degree in Aerospace Engineering from ISAE-ENSMA (France), an M.Sc. degree in Space Instrumentation and his Ph.D from Paris Observatory. He is the Technical Director of CENSUS \cite{census} , space pole of PSL Université Paris dedicated to scientific nanosatellites.
\end{biographywithpic} 

\begin{biographywithpic}
{Youssoupha Diaw}{illustrations/youss_resized}
was a student of the Graduate Program AstroParis by PSL Université Paris (France), intern at CENSUS \cite{census}, at the time of his contribution in 2021 for this paper. He is now graduating to a Master's degree in Space Instrumentation at Paris Observatory.
\end{biographywithpic}

\begin{biographywithpic}
{Valéry Lainey}{illustrations/VLainey}
defended his Ph.D in 2002 on orbital dynamics of the Galilean moon system. He was Deputy Coordinator of the FP7-ESPaCE network and the coordinator of the ENCELADE international team. Recently, he spent two years at JPL (Caltech, USA) working on Cassini data.
\end{biographywithpic}

\end{document}